\documentstyle[12pt,epsf]{article} 
\newcommand{\preprint}[1]{\begin{table}[t]              
\begin{flushright}                          
\begin{large}{#1}
\end{large}                
\end{flushright}                            
\end {table}} 
\preprint{TAUP-2414-97} 
\tiny\normalsize

\begin{document} 
\title{The Associated Metric for a Particle in a Quantum Energy Level.} 
\author{E. Atzmon\thanks{atzmon@post.tau.ac.il}  \\
Raymond and Beverly Sackler Faculty of Exact Sciences,\\ School of Physics 
and Astronomy.\\Tel\ - Aviv University. } 
\date{\today} 
\maketitle 
 
\begin{abstract} 
We show that the probabilistic distribution over the space in the spectator 
world, can be associated via noncommutative geometry (with some
modifications) to a metric in which the particle lives. According to
this geometrical view, the metric in the particle world is
``contracted'' or ``stretched'' in an inverse proportion to the
probability distribution.
\end{abstract} 
 
\newpage 
 
\section{Prelude} 
 
It is well known that for many physical quantum potentials, the spectrum of 
the Hamiltonian is a discrete set (i.e. the set of energy
eigenvalues $E_n$ is a countable set) associated with a separable
Hilbert space of eigenstates $\psi _n$.\footnote{Over compact spaces the spectrum of the Hamiltonian is separable.} It is also well known that in Quantum Mechanics, $\left| \psi (x)\right| ^2$ defines a probability 
distribution over coordinate space, in the observer world. Using the formula 
for distance in noncommutative geometry \cite{con}, we were recently able to 
show that one might then also associate the relevant metric with particle 
energy levels. 
In noncommutative geometry, one uses the spectral triple $(A,H,D)$ , where $%
A $ is the algebra of functions on the space (which according to the 
Gelfand-Naimark theorem \cite{gn} is fully characteristic of the space), $H$ 
is a Hilbert space, and $D$ is a Dirac operator, which acts as a linear 
operator on that Hilbert space. The Dirac operator is a self-adjoint 
operator, with compact resolvent. In recent years,noncommutative
geometry \cite{models} has provided the means of further extending the
geometric approach to gauge theories with spontaneous symmetry
breakdown. In this study, we point out yet another possible
application, namely the definition of a metric associated with quantum
spectra. At this stage we limit our treatment to its evaluation and a
discussion of its consistency with the probabilistic intuitive picture.\\
  
The distance formula in noncommutative geometry is:  
\begin{equation} 
\label{dis}d(a,b)=\sup _f\left\{ \left| f(a)-f(b)\right| \;:\;f\in 
A,\;\left\| \;\left[ \;D,f\;\right] \;\right\| \leq 1\right\}  
\end{equation} \\ 
where $a,b\in X,\;f\in A,$\ $A$ is the algebra of functions on $X,$ and 
the norm on the r.h.s is the norm of operators in $H$. Both the 
noncommutative geometrical and the classical (i.e. the infimum among all the 
paths which connect $a$ and $b$) definitions for distance give the same 
result when the base space is a Riemannian manifold of genus zero. However 
the n.c.g definition has the advantage of being applicable to discrete 
spaces too, on which the concept of a ``path'' is not well defined.\\ 

The main advantage of the distance formula, eq.(\ref{dis}), is that it
involves not just the algebra of functions defined on the space, but
also the Hilbert 
space, which is usually associated with Quantum Mechanics. Had 
the geometry's defining triple not involved a Hilbert space of states, the 
algebra of functions would have been applied to a fixed manifold (i.e. a 
topological manifold, without any dynamics). However, once we have a 
Hamiltonian $H$, its spectrum is associcated with a Hilbert space of
eigenstates. The eigenstates essentially fix the probability
distribution of the particle's position in space.\footnote{In Quantum
  Mechanics the probability distribution of the particle's position in
  space applies to a single particle, and not only as a statistical
  result due to large number of measurements. This is just as in the
  two slits experiment, where a single particle passes through both slits.} The base space can thus now be considered as a dynamical space, i.e., a space depending on the quantum state (and therefore also on the energy level). This can be seen in every scattering process, e.g. note how visible 
light is scattered on drops forming a rainbow, due to spectral 
decomposition, so that each frequency has it own geodesic line. One might 
well claim that there is an interplay between the fixed base space, and the Hilbert space.\\ 

In the following section we will give a metric interpretation, as
``seen'' by the particle, in its own world, which is consistent with
the distributive aspects assigned to the wave function, in quantum
mechanics for the observable world.\\

\section{The Quantum Mechanical Distance Formula} 
Let assume for simplicity that we are giving a one dimensional
Hamiltonian system which found in its lowest eigenstate $\psi$. While using eq.(\ref{dis}) over a smooth manifold, one can prove that:  
\begin{equation} 
\label{sup}\sup _f\left| f(a)-f(b)\right| =\sup _f\left| \int_a^b\left| 
\nabla f\right| ds\right| =\left| \int_a^b\sup _f\left| \nabla f\right| 
ds\right|  
\end{equation}
(where a, b not necessarily ordered).\\
The norm  in eq.(\ref{dis}) usually defined as follows:
\begin{equation} 
\label{normst}\left\| \;\left[ \;D,f\;\right] \;\right\| = \sup
\left\{\left\| \;\left[ \;D,f\;\right]\;\psi\;\right\| \;\;\;\;\;
  \mbox{where}\;\;\;\left\|\psi\right\|\le 1\right\} 
\end{equation}
However, we suggest to modify the norm condition in eq.(\ref{dis}) to be
interpreted as a condition on the expectation value of
$\left|\left[\;D,f\;\right]\right|$ in the quantum state $\psi$ - that is:  
\begin{equation} 
\label{norm}1\ge \left\langle\;\left|\left[ \;D,f\;\right]\right|\;\right\rangle = 
\left\langle \psi \left| \left| \left[ \;D,f\;\right] \right| \right| \psi 
\right\rangle = \int_v\psi ^{*}(x)\left| \left[ \;D,f\;\right] \right| \psi (x)dx  
\end{equation} 
where $v$ is the entire base space, and we are assuming also that
$\psi$ is normalized:  
\begin{equation} 
\label{nirmul}\left\langle \psi |\psi \right\rangle =\int_v\psi ^{*}(x)\psi 
(x)dx=1  
\end{equation} 
Therefore, in order to be consistent with both eq.(\ref {nirmul}) and
the norm condition in eq.(\ref{dis}) replaced by the new ``norm''
condition in eq.(\ref{norm}), it is sufficient to require the following
local condition:  
\begin{equation} 
\label{jakob}\left| \left[ \;D,f\;\right]
\right|\left.\right|_{x} =\left| \nabla f\right|\left.\right|_{x}\leq \frac 1{Vol(v)\psi ^{*}(x)\psi (x)}  
\end{equation}
where $Vol(v)$ is the world volume that we assume to be finite. The
idea to replace the norm condition in eq.(\ref{dis}) by eq.(\ref{norm}),
follows from the fact that in Q.M. one considers only expectation values as
observables. This would enable a quantum state to affect the observed
metric as will be shown immediately.  
By applying eq.(\ref{jakob}) to eq.(\ref{sup}), one should use the 
equality in eq.(\ref{jakob}). So if $\forall x\in X,\exists \psi (x)\not =0$ 
(except on the boundaries), which is always the situation at the first 
energy level, the distance formula eq.(\ref{dis}) becomes:  
\begin{equation} 
\label{no0}d(a,b)=\frac 1{Vol(v)}\left| \int_a^b\frac 1{\psi ^{*}(x)\psi 
(x)}ds\right|  
\end{equation} 
where $ds$ is a path element on the straight line which connects the points $%
a$ and $b$.\\ 

Assume now that we are working in a $D$-dimensional base 
space, such that the wave function of a particle which is found at some 
energy level is given by: $\psi (x_1,...,x_D)=\prod_{i=1}^D\psi _{n_i}(x_i)$%
, where the $i$-index labels the coordinates, and the $n_i$-index indicates 
in which eigenstate the particle is found (according to zero locus of
$\psi$), over the $i$-coordinate (i.e. we assume separability of the 
variables). The base space is thus being 
divided naturally by the distribution probability, defined on the base 
space, into $N\equiv \prod_{i=1}^Dn_i$ parts (induced by the zero
locus of $\psi$). One should thus consider the 
particle as being situated in any of the N parts, i.e. the location of the 
particle is now being represented by an $N$-dimensional vector. Therefore, 
one considers the particle as situated at one spatial point, given in $N$%
-dimensional space in its own world, but in $D$-dimensions in the outside 
world.\\ \\ 
\vspace{10mm} \epsfxsize=8truecm \centerline{\epsfbox{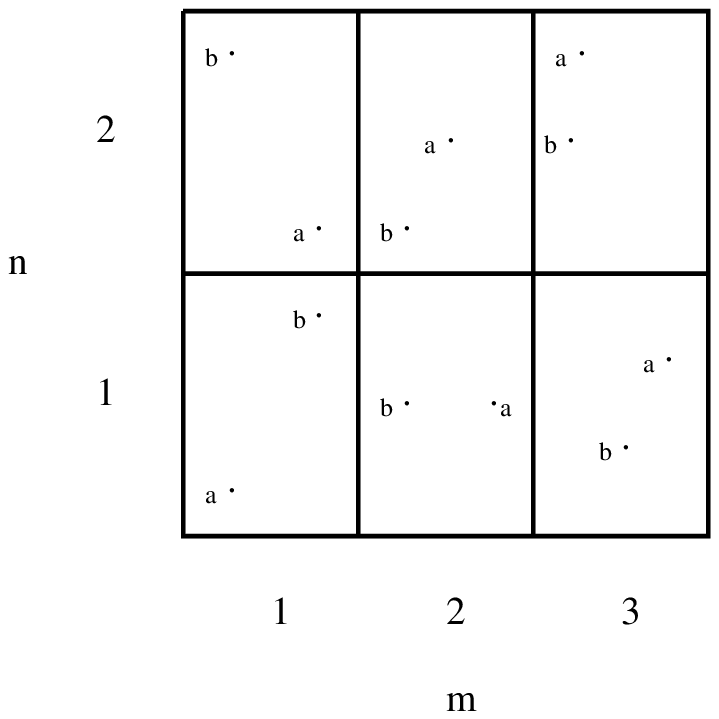}}  
\vspace{5mm}  
\centerline{\parbox{13truecm}{Figure 1. {\footnotesize Two points 
      $a\;,\;b$ in the particles` world, while the particle is found in 
      a two dimensional infinite square well potential at the energy 
      level $n=2,\; m=3$. So $N=6$ (since there are 6 cells) and $D=2$ and therefore in this case its location will be defined by a $6\;\times\;2$ matrix}}} 
\\ \\ The full representation of the location of the particle thus requires 
an $N\times D$ - matrix, where each row represents a domain. The distance 
formula will thus measure the distance between two points in $N$-dimensional 
space, treated as a vector space, i.e., weighted in a Pythagorean way, and 
inside each domain as in eq.(\ref{no0}). The quantum distance formula now 
takes the following form:  
\begin{equation} 
\label{e0}d_N(a,b)=\sqrt{\sum_{i=1}^N\left( \frac 1{Vol(v_i)}\left| 
\int_{a_i}^{b_i}\frac 1{\psi ^{*}(x)\psi (x)}ds_i\right| \right) ^2}  
\end{equation} 
where the $i$-index runs over the domains (i.e. over the rows), and by
$Vol(v_i)$ one means:
\begin{equation}\label{volvi}
Vol\left(v_i\right) = Vol(v)\int_{v_i}\psi ^{*}(x)\psi (x)dx
\end{equation}\\ 
In order that eq.(\ref{e0}) would be applied also to spaces with
infinite volume we suggest to replace the definition of $Vol(v)$. We
choose either to define $Vol(v)$ as the classical volume\footnote{For
  example: for harmonic oscillator one can take the volume as based on
  the amplitude of the corresponding energy level, or for the hydrogen
  atom one can take the appropriate Bohr radius (see examples 3.3, 3.4).} or to
define $\frac{1}{Vol(v)}$ as the square of the normalization
pre-factor appears in the eigenstate (up to an arbitrary scale).\\

In cases where the particle state is given by a superposition of
eigenstates, one has to find the $(D-1)\;-$ subspace on which the wave
function vanishes. The space is thereby naturally separated into
parts. As explained, the particle 
is represented in each of these parts. The $D\;-$ dimensional volume of each 
part should also be evaluated, using eq.(\ref{e0}). However, the main 
difference is that in this case the metric evolves with time. Each 
eigenstate in the superposition has its own time evolution, therefore the 
probability distribution $\left| \psi \right| ^2$ also depends on time 
(whereas there is no such time-dependence in the case in which the particle 
is found in an eigenstate).\\
 
In the following section we apply the quantum 
distance formula eq.(\ref{e0}) to several examples and discuss some of the 
implications of the results. 
 
\section{Examples} 
 
\subsection{One Dimensional - Infinite Square Well Potential} 
 
The potential is defined as:  
\begin{equation} 
\label{pot1}U\left( x\right) =\left\{  
\begin{array}{cc} 
0 & \;\;\;0\le x\le L \\  
\infty & \;\;\;\mbox{elsewhere}  
\end{array} 
\right.  
\end{equation} 
\\ So, the eigenstates are: $\psi _n(x)=\sqrt{\frac 2L}\sin \left( \frac{%
n\pi x}L\right) $ and $Vol(v)=L$. The distances at the first energy
level are therefore,  
\begin{equation} 
d_1(a,b)=\frac 1L\left| \int_a^b\frac 1{\frac 2L\sin ^2\left( \frac{\pi x}%
L\right) }dx\right| =\frac L{2\pi }\left| \cot \left( \frac{\pi b}L\right) 
-\cot \left( \frac{\pi a}L\right) \right|  
\end{equation} 
\\ \vspace{0mm} \epsfxsize=15truecm \centerline{\epsfbox{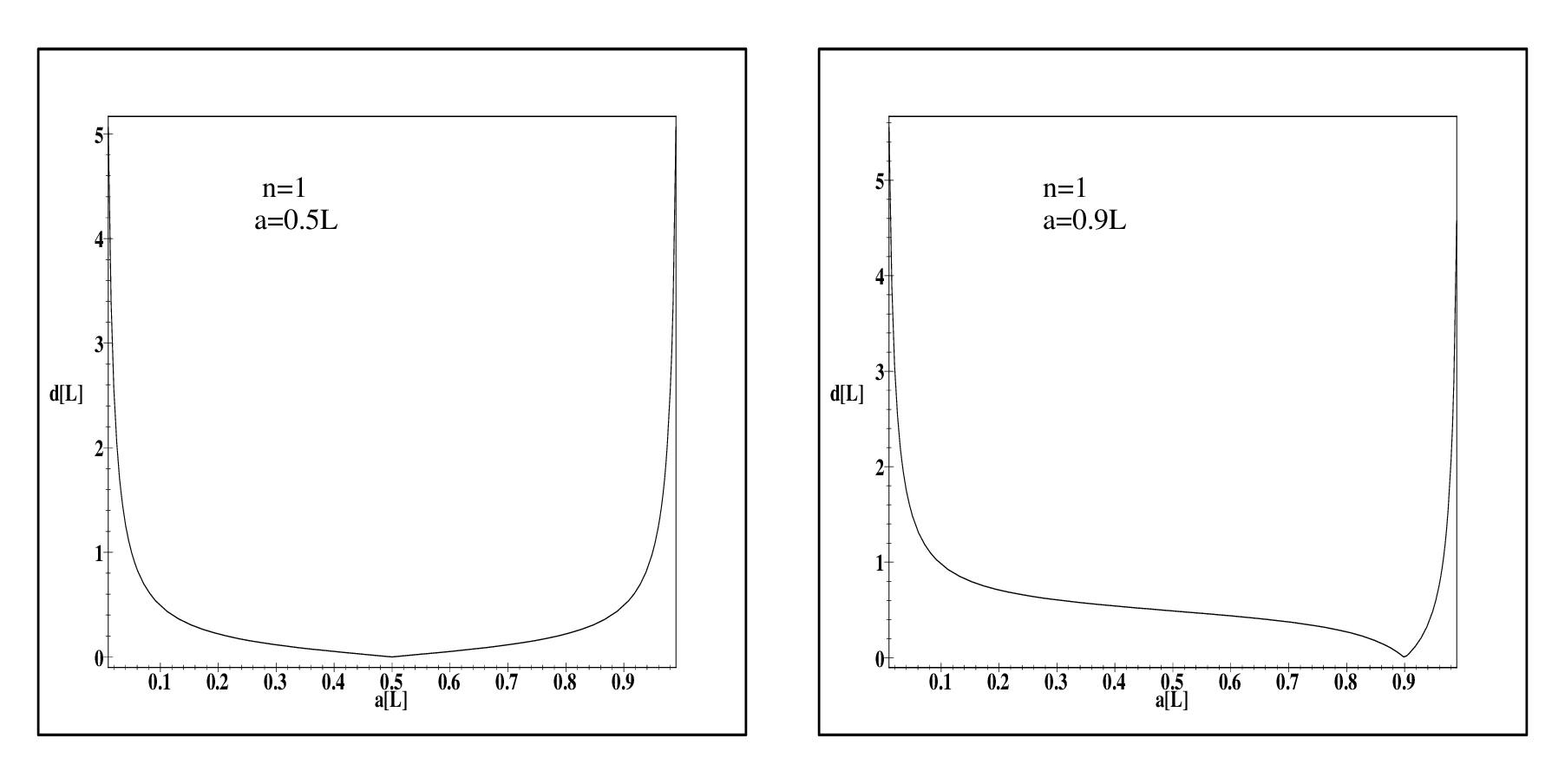}}  
\vspace{5mm}  
\centerline{\parbox{13truecm}{Figure 2.{\footnotesize  The distances in 
      an infinite square well potential, at the first energy level, as 
      seen from the points a=0.5L and a=0.9L}}} \\ \\ The distance at the $%
n^{th}$- level then becomes:  
\begin{equation} 
d_n(a,b)=\frac {n}{L}\sqrt{\sum_{i=1}^n\left| \int_{a_i}^{b_i}\frac 1{\frac 
2L\sin ^2\left( \frac{n\pi x}L\right) }dx\right| ^2}=\frac L{2\pi }\sqrt{%
\sum_{i=1}^n\left| \cot \left( \frac{n\pi b_i}L\right) -\cot \left( \frac{%
n\pi a_i}L\right) \right| ^2}  
\end{equation} 
where $a_i,b_i\;\in \;\left( \frac{(i-1)}nL\;,\;\frac inL\right) $ and $%
i=1,2,...,n$.\\ 
 
\subsection{Two Dimensional - Infinite Square Well Potential} 
 
The potential is defined as follows:  
\begin{equation} 
\label{pot2}U\left( x,y\right) =\left\{  
\begin{array}{ccc} 
0 & \;\;\;0\le x\le L &  \\  
0 & \;\;\;0\le y\le L &  \\  
\infty & \;\;\;\mbox{else where} &   
\end{array} 
\right.  
\end{equation} 
\\ The eigenstates are then: $\psi _{n,m}(x)=\frac 2L\sin \left( \frac{n\pi x%
}L\right) \sin \left( \frac{m\pi y}L\right) $\\ The distance formula then 
becomes:  
\begin{equation} 
\label{ddis}d_{\left( n,m\right) }\left( a,b\right) =\frac{nm}{L^2}\sqrt{%
\sum_{i=1}^{nm}\left| \int_{a_i}^{b_i}\frac 1{\frac 4{L^2}\sin ^2\left(  
\frac{n\pi x}L\right) \sin ^2\left( \frac{m\pi y}L\right) }ds_i\right| ^2}  
\end{equation} 
where the $i-$index stands for the $i^{th}-$zone (where there are $nm$ 
zones), and $a_i\;,\;b_i$ are respectively the initial and final points in 
the $i^{th}-$zone.\\ 
 
\subsection{One Dimensional Harmonic Oscillator} 
 
The potential is:  
\begin{equation} 
\label{pot3}U\left( x\right) =\frac 12m\omega ^2x^2  
\end{equation} 
\\ The eigenfunctions are $Hermite\;polinomials$:  
\begin{equation} 
\label{herm}\phi _n(x)=\left[ \frac 1{2^nn!}\left( \frac \hbar {m\omega 
}\right) ^n\right] ^{1/2}\left( \frac{m\omega }{\pi \hbar }\right) 
^{1/4}\left[ \frac{m\omega }\hbar x-\frac d{dx}\right] ^ne^{-\frac 12\frac{%
m\omega }\hbar x^2}  
\end{equation} 
\\ so that the $0^{\;th}$ - eigenstate is:  
\begin{equation} 
\label{phi0}\phi _0(x)=\left( \frac{m\omega }{\pi \hbar }\right) 
^{1/4}e^{-\frac 12\frac{m\omega }\hbar x^2}  
\end{equation} 
As the volume we take the ``classical'' amplitude $A_0$ at this energy
level which can be found by equating the energy of this quantum level with the potential energy - i.e. $%
\frac 12m\omega ^2A_0^2=\frac 12\hbar \omega $. Thus, the ``classical'' 
amplitude at this energy level is $A_0=\sqrt{\frac \hbar {m\omega}}$. 
Therefore the distance at the $0^{\;th}\;energy-level$ is:  
\begin{eqnarray} 
\label{dis3}d_0\left(a,b\right)=\int_{a}^{b}\sqrt{\pi}e^{\frac 
  {m\omega}{\hbar}x^2}dx = \left(\frac{\pi}{2}\sqrt{\frac 
    {\hbar}{m\omega}}\right)\left|\left[{\mbox erf} 
    \left(i\sqrt{\frac{m\omega}{\hbar}}b\right)-{\mbox erf} 
    \left(i\sqrt{\frac{m\omega}{\hbar}}a\right)\right]\right|=\\ 
\frac{\pi}{2} A_0 \left|\left[{\mbox erf}\left(i \; \frac{b}{A_0}\right)-{\mbox erf}\left(i\; \frac a{A_0}\right)\right] \right| \nonumber 
\end{eqnarray}\\
For higher energy levels one has to consider the 
``classical'' amplitudes $A_n=A_0\sqrt{2n+1}$ as the total volume - $%
Vol\left( v\right) $. One has then to find the $n-$roots of the $n^{\;th}-$%
Hermite polynomial. Denote the $n-$roots by $\left( \alpha _1,...,\alpha 
_n\right) $. The partial volumes $Vol\left( v_i\right) $ in eq.(\ref{e0}) 
become $Vol\left( v_i\right) =Vol\left( v\right) \int_{\alpha 
_{i-1}}^{\alpha _i}\left| \phi _n\left( x\right) \right| ^2dx\;$ where $%
\;\alpha _0=-\infty \;$ and $\;\alpha _{n+1}=\infty $. One can now make use 
of eq.(\ref{e0}) to find the distances at higher energy
levels. Evaluating the distance associated with the harmonic oscillator at higher dimensions is a straightforward calculation, making use of eq.(\ref{e0}).\\ 
\vspace{0mm} \epsfxsize=15truecm  
\centerline{\epsfbox{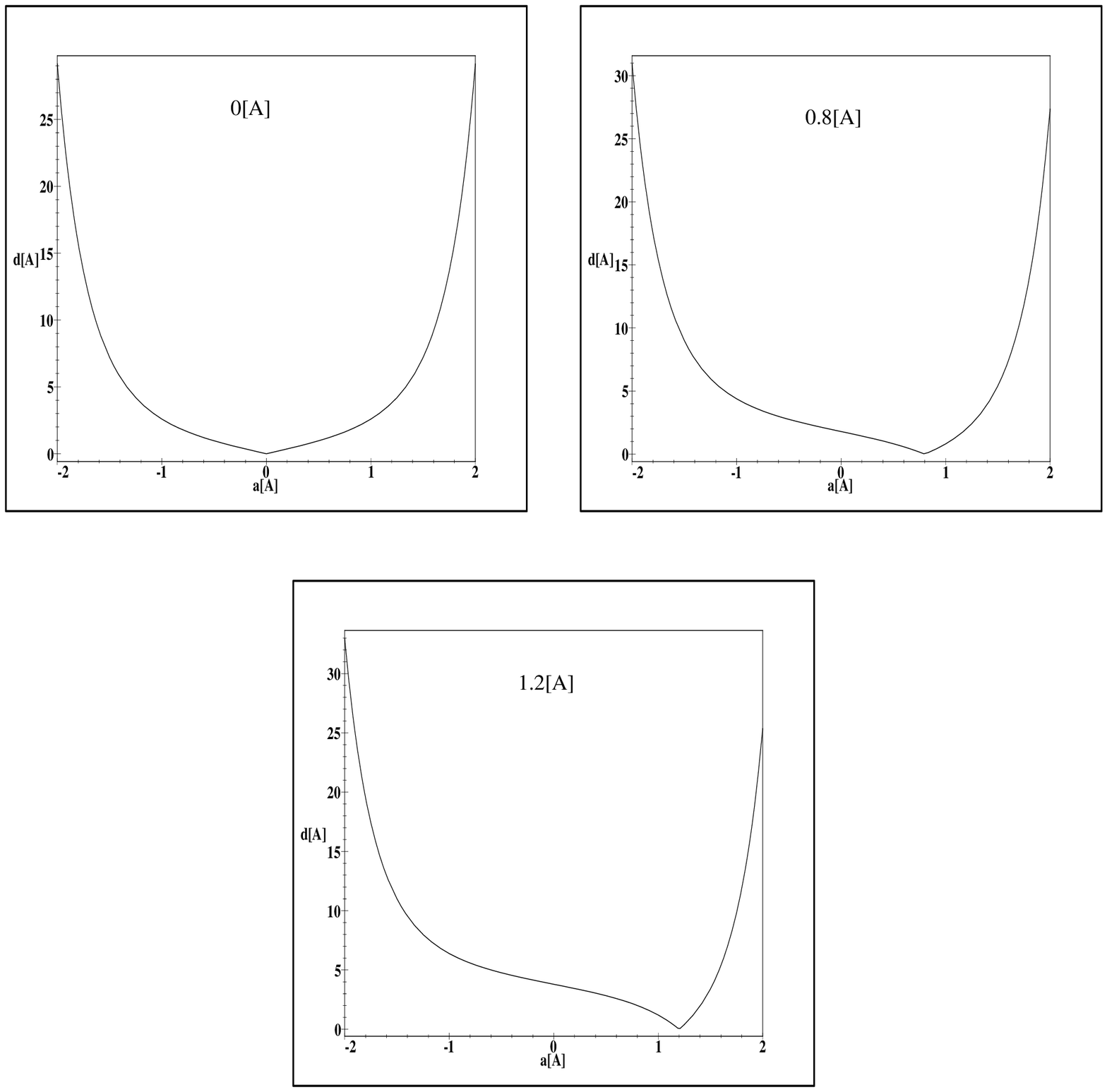}} \vspace{5mm}  
\centerline{\parbox{13truecm}{Figure 3.{\footnotesize  The distances in 
      a one dimensional harmonic oscillator, at the n=0 energy level, as seen from the points a=0[A], a=0.8[A] and a=1.2[A]}}} 
\vspace{5mm}\\ 

\subsection{The Hydrogen Atom} 
 
The potential is given by:  
\begin{equation} 
V(r)=-\frac{e^2}r  
\end{equation} 
For simplicity, we compute the distances, as seen in the electron's world, 
in the $1s$ and the $2s$ levels and take only distances between two points 
situated on the same radius vector.\\ The $1s$ and the $2s$ eigenstate are:  
\begin{equation} 
\begin{array}{l} 
1s:\,\,\,\,\,\,\,\,\,\phi _{1s}=\frac 1{ 
\sqrt{\pi a_0^3}}\,e^{-r/a_0}\\  
2s:\,\,\,\,\,\,\,\,\,\phi _{2s}=\frac 1{ 
\sqrt{8\pi a_0^3}}\,\left( 1-\frac r{2a_0}\right) \,e^{-r/(2a_0)}  
\end{array} 
\end{equation} 
where $a_0$ is the Bohr radius. In the $1s$ case we define $Vol\left( 
v\right) =\frac{4\pi }3a_0^3$\thinspace \thinspace \thinspace and the 
distance between two points along the same radius vector is:  
\begin{equation} 
d_{1s}\left( r_1,r_2\right) =\frac 38\,\left| {\mbox e}^{2r_2/a_0}-{\mbox e}%
^{2r_1/a_0}\right| \,a_0  
\end{equation} 
In the $2s$ case one sees that the eigenstate vanishes at $r=2a_0$ and the 
space is thereby naturally split into two domains: $r<2a_0$ and $r>2a_0$. 
This means that according to our definition the electron's location is 
represented in both sectors - \\ i.e. $\left\{ {}\right. \left( 
x_1,x_2\right) \,\,\,\,\,$where$\,\,0\le x_1<2a_0\,\,\,\,\,$and$%
\,\,2a_0<x_2<\infty \left. {}\right\} $. The inner and outer volumes, 
appearing in eq.(\ref{e0}), will be:  
\begin{equation} 
\begin{array}{l} 
Vol\left( v_1\right) =  
\frac{4\pi }3(R_2)^3\int_0^{2a_0}\left| \phi _{2s}\left( r\right) \right| 
^24\pi r^2dr=\frac{256}3\pi a_0^3\;[1-\frac 7{{\mbox e}^2}] \\ Vol\left( 
v_2\right) =\frac{4\pi }3(R_2)^3\int_{2a_0}^\infty \left| \phi _{2s}\left( 
r\right) \right| ^24\pi r^2dr=\frac{256}3\pi a_0^3\;\frac 7{{\mbox e}^2}  
\end{array} 
\end{equation} 
where we make use of the Bohr radius of the n$-th$ level, defined as: $%
R_n=n^2a_0$. One can now make use of eq.(\ref{e0}) and the distance between 
the point $(x_1,x_2)$ and the point $(x_1^{\prime },x_2^{\prime })$ becomes:  
\begin{equation} 
d_{\left( x_1,x_2\right) }^{\left( x_1^{\prime },x_2^{\prime }\right) }=%
\sqrt{\left( \frac 1{Vol\left( v_1\right) }\left| \int_{x_1}^{x_1^{\prime 
}}\left| \phi _{2s}\right| ^{-2}dr\right| \right) ^2+\left( \frac 
1{Vol\left( v_2\right) }\left| \int_{x_2}^{x_2^{\prime }}\left| \phi 
_{2s}\right| ^{-2}dr\right| \right) ^2}  
\end{equation} 
\section{Discussion and Conclusions} 
It is worth reviewing 
the distances evaluated in the infinite square well potential, as compared 
to those reached in the harmonic oscillator case. While in an infinite 
square well potential the distances at each energy level are independent of 
both the mass of the particle and $\hbar $, in the harmonic oscillator the 
distances depend on the mass of the particle and on the oscillator frequency 
- and also on $\hbar $. The fact that $\hbar $ does not appear in the 
distance formula in the infinite square well potential case might hint at 
one of the following:\\ 1. Such potentials do not really exist in nature, 
but can be constructed as a limiting case of a superposition of harmonic 
oscillators.\footnote{%
There should be an infinite of them, since the energy spectrum of the
infinite square well potential is proportional to $n^2$ while for the
harmonic oscillator it is proportional to $n$.} 
In that case the walls of the infinite square well 
potential should be associated with the ``classical amplitudes'' of the 
harmonic oscillator (i.e. $A_i=\sqrt{\frac \hbar {m_iw_i}}$ where the $i-$ 
index label the harmonic oscillators - see also eq.(\ref{dis3})\,).\\ 
2. On the 
other hand, if infinite square well potentials do exist in nature, then a 
spectrum of $mw$ should be induced, due to our former reasoning.\\

As one can see the distance between any point $x$ 
for which $\psi (x)\neq 0$ and a point $y$ for which $\psi (y)=0$ becomes 
infinite. The less the probability distribution in the outside world, the 
larger the distances in the particle world at that neighborhood, i.e. the 
metric is `stretched'. In other words, while $\left| \psi (x)\right| ^2$ is 
the probability distribution as defined over the space in the outside world,  
$\frac 1{\left| \psi (x)\right| ^2}$ is the distribution of points as seen 
within the particle world. This can conceptually be understood from the 
similarity to the W.K.B. interpretation.\\ In the W.K.B approximation $\frac 
1{\left| \psi \left( x\right) \right| ^2}$ is interpreted as the average 
velocity of the particle in space. The velocity is however responsible for 
the particle's translation in space. Thus, the velocity can be interpreted 
as a particle's passage through a density of points in space. The larger the 
velocity, the larger the spatial point-density. The differential element $ds$ 
in the quantum distance formula should therefore be associated with time, 
for the quantum formula to have a metrical meaning. Thus, in the particle 
world, ``farther'' means (according to the W.K.B approximation) a longer 
time in the spectator's outside world. That is to say that the spectator 
would have to wait a long time until he might find the particle at a place 
with a low probability density.\footnote{%
Note that the interplay between time and distance elements already
occurs when, one compares the metric outside and inside a black
hole. It seems not to be an accident, since in both cases we have a compact space which is 
disconnected from the observer.}\\ 

Another fact one should be aware of is 
that the natural splitting induced by the probability distribution $\left| 
\psi \left( x\right) \right| ^2$ over space, in the spectator world, is 
being translated in the particle world into the fixing of the dimension of 
the space. The above conclusions can be observed in all the examples
given. In this manner, the higher the energy level, the higher is the
space dimensionality in the particle world.\\
 
One might ask - is the metric (in the particle world) detectable?! We might 
claim that any measuring process would insert a new interaction potential 
into the Hamiltonian, and that this would modify the eigenstates and thereby 
change the metric. Moreover, due to the effect of the turning on of that 
interaction potential, the change in the metric would be time dependent. 
This is just one more case in Quantum Mechanics, in which the measuring 
process affects the system which is being measured. However the question is 
how drastic is that effect. We claim that it might sometime be quite 
drastic, if the new eigenstate of the system and of the measurement device 
be changed in such a way that the quantum level is modified. This would 
cause a topological change - i.e. a change in the dimension (due to the 
change in the number of zones, as defined by the probability distribution).%
\footnote{%
It is well known that in the two slits experiment, if one tries to detect 
the ``geodesics (i.e the metric)'' a change in the topology is induced.} 
However, if the energy level is slightly shifted - i.e. in such a way that 
the subspace on which $\psi \left( x\right)=0 $ changes smoothly, then the change in the metric will be a minor 
one.\footnote{%
The same effect exists classically - since in any process of measuring
a metric, at the end, the information should be sent back to the 
observer. However, the information is being carried by some particle which 
has some energy. Thus, $T_{00}$ is reduced and with it the metric is 
changed. If the observer himself wants to be part of the metric, then by 
that he is changing the $T_{00}$ and with it the endowed metric.}\\ 
\newpage
{\large Acknowledgment}\\
I'm most greatful to Prof. Y. Ne'eman and Prof. L. P. Horwitz for useful
discussions. I am greatful to Prof. A. Connes for inviting me to the IHES, where this work was partially done, and to the IHES for its warm hospitality.\\

\end{document}